\documentclass{iaus}
\usepackage{graphicx}
\usepackage{natbib}

\title[The multiplicity of massive stars] %% give here short title %%
{The multiplicity of massive stars}

\author[H. Sana \& C. J. Evans]   %% give here short author list %%
{Hugues Sana$^1$
 \and Christopher J. Evans$^2$}

\affiliation{$^1$Sterrenkundig Instituut Anton Pannekoek,
 University of Amsterdam, \\ Postbus 94249, NL-1090~GE Amsterdam,
 The Netherlands \\ email: {\tt h.sana@uva.nl} \\[\affilskip]
$^2$UK Astronomy Technology Centre, Royal Observatory Edinburgh \\
Blackford Hill, Edinburgh, EH9 3HJ, UK \\ email: {\tt chris.evans@stfc.ac.uk }}

\pubyear{2011}
\volume{272}  %% insert here IAU Symposium No.
\pagerange{1-12}
\jname{Active OB stars}
\editors{C. Neiner, G. Wade, G. Meynet \& G. Peters, eds.}

\begin{document}

\maketitle

\begin{abstract} 
Binaries are excellent astrophysical laboratories that provide us
with direct measurements of fundamental stellar parameters. 
Compared to single isolated star, multiplicity induces new processes,
 offering the opportunity to confront our understanding of a broad range 
of physics under the extreme conditions found in, and close to, astrophysical objects. 

In this contribution, we will discuss the parameter
space occupied by massive binaries, and the observational means to
investigate it.  We will review the multiplicity fraction of OB stars 
within each regime, and in different astrophysical environments. 
In particular we will compare the O star spectroscopic binary fraction in 
nearby open clusters and we will show that the current data are adequately described by an homogeneous fraction of $f\approx0.44$. 

We will also summarize our current understanding of the observed parameter distributions
of O\,$+$\,OB spectroscopic binaries. We will show that the period distribution is overabundant in short period binaries and that it can be described by a bi-modal \"{O}pik law with a break point around $P\approx 10$~d. The distribution of the mass-ratios shows no indication for a twin population of equal mass binaries and seems rather uniform in the range $0.2\le q=M_2/M_1\le1.0$.

\end{abstract}

\firstsection % if your document starts with a section,
              % remove some space above using this command.

\section{Introduction}\label{sect: intro}

Massive stars have many fascinating aspects, which extend well
beyond stellar physics alone. One of their most striking
properties is conceptually very simple: their 
high-degree of multiplicity. Most O- and early B-type stars are found
in binaries and multiple systems. Even single field stars are often
believed to have been part of a multiple system in the past, then ejected
by a supernova kick or by dynamical interaction.  To ignore the
multiplicity of early-type stars is equivalent to neglecting one of
their most defining characteristics.

In this review we concern ourselves with the multiplicity of stars
more massive than 8~M$_\odot$ on the zero-age main sequence, which
have spectral types earlier than B3~V.  Our approach is to focus on their
observational properties, with the emphasis on O-type binaries,
although early B-type binaries feature in some of the quoted
works.  Despite the importance of detailed studies of individual
objects, our prime motivation here is to consider the broader results
from the literature, in an attempt to lift the veil on
some of the general properties of the binary population of early-type
stars.

The distributions of the orbital parameters of massive binaries, as a
population, are of fundamental importance to stellar evolution, yet
remain poorly constrained. These distributions trace the products of
star formation and the early dynamical evolution of the host systems,
and are necessary ingredients to population synthesis studies. Only
with an understanding of these distributions can we hope to recover
accurate predictions for some of the exotic late stages of binary
evolution.

This contribution is structured as follows. Section~\ref{sect: physic}
describes some of the physical processes and observational biases that
are present in multiple systems compared to single stars.
Section~\ref{sect: param} introduces the different parts of parameter
space occupied by massive binaries, and the observational means to
investigate them; Section~\ref{sect: fbin} then reviews the
multiplicity fraction of OB stars within each regime, and in
different astrophysical environments. Section~\ref{sect: CDF} attempts
to summarize our current understanding of the parameter distributions
of O\,$+$\,OB spectroscopic binaries. Finally, Section \ref{sect: ccl}
provides a summary.

\section{Physical processes and observational biases}\label{sect: physic}

Binaries are excellent astrophysical laboratories that provide us
with direct measurements of fundamental parameters such as stellar
masses and radii. Multiplicity induces new processes compared to
isolated single stars, offering the opportunity to confront our
understanding of a broad range of physics under the extreme conditions
found in, and close to, astrophysical objects. Moreover, if one fails to take
multiplicity into account, observations (and their analysis) can be significantly
biased or misleading. Most critically, early-type binaries with
orbital periods of up to 10 years follow significantly
different evolutionary paths, an aspect that can also impact the
outputs of population synthesis models \citep[e.g.,][]{Van09}. By way of additional motivation to
understand multiplicity in massive stars, some of the observational
and evolutionary impacts include:

{\underline{\it Different evolutionary paths:}} Binarity significantly affects
the evolutionary path of the components of the systems compared to
single stars. Tidal effects in close binaries modifies the evolution of
stellar rotation rates, thus also the induced rotational-mixing of
enriched material into their photospheres \citep{dMCL09}. Roche-lobe overflow will
result in mass and angular momentum transfer, spinning up the
secondary to its critical rotation rate \citep{Pac81,LCY08}. While the gaining star might
be rejuvenated by the increase in mass \citep{BrL95}, the primary will see a
reduction in the life-time of its red supergiant phase \citep{EIT08}. A
common-envelope phase and/or stellar mergers are other possible
outcomes of binary evolution. The impacts on observed stellar
populations are numerous, including modified surface abundances,
modified enrichment of the interstellar medium, the rate of supernova
 and $\gamma$-ray burst explosions, and on the number of
evolved systems such as Wolf-Rayet stars and high-mass X-ray binaries
\citep[e.g.,][]{IDK06,BKD08}.

{\underline{\it Wind collisions:}} In binaries, the powerful
stellar wind from the stars may interact with one another or with the
surface of the star with the weaker wind \citep{Uso92}. The supersonic
collision heats the gas to temperature up to several 10$^7$~K
\citep{SBP92}. In several cases, the wind-wind interaction is also to
 accelerate particles up to relativistic energies. The
signature of the wind collision can be observed throughout the
electromagnetic spectrum, through non-thermal radio (and possibly X-
and $\gamma$-ray) emission \citep{DeB07}, through X-ray thermal
emission \citep{PiP10} and via a contribution to the recombination lines
in the optical and infrared \citep{SRG01}. In massive binaries containing 
evolved stars with very dense winds, the wind interaction region can act 
as a nucleation site for dust particles, creating structures such as the 
pinwheel nebulae \citep{TML08}. These effects can provide
indirect indiciations of multiplicity.  However, if multiplicity is
not considered, wind collision can lead to erroneous estimates of fundamental
properties such as intrinsic X-ray luminosities \citep{SRN06},
spectral classifications, and stellar mass-loss rates (as measured
from the strength of, e.g., the H$\alpha$ line).

{\underline{\it Struve-Sahade effect:}} In its most generalized form,
the Struve-Sahade (S-S) effect can be described as the variation in
the apparent strength of the spectrum of one or both components when
the star is approaching/receding \citep[for an example, see e.g. ][]{SRG01}. Various physical effects can induce
a S-S signature: gaseous streams in the systems, ellipsoidal
variations, surface streams, and changes in the local surface
temperature due to, e.g., mutual illumination or heating from a
wind-wind collision \citep[e.g.,][]{BGR99, LRS07}. 

{\underline{\it Cluster dynamical mass:}} Ignoring the contribution of
binaries to the stellar velocity dispersion in clusters (in both
integrated-light observations of distant systems and studies of
resolved clusters), can lead to a significant overestimate of their
dynamical mass \citep{BTT09,GSPZ10}. For example, some of the
disagreement in the mass-to-light ratio of young extragalactic
clusters might arise from the binary properties of their red
supergiant populations \citep{GSPZ10}.

{\underline{\it Supermassive stars:}} Unresolved multiple systems have
often been confused with very high mass stars due to their large
luminosity. Numerous objects have indeed seen their masses revised at
the light of improvements of the observing facilities \citep[e.g. the case of R136:][]{CMS81, WeB85, CSH10}. 

\section{The parameter space}\label{sect: param}
Before discussing the multiplicity properties of populations of
massive stars, we attempt to give the reader a feel for
the typical parameter space that needs to be investigated. Our aim
is to provide a qualitative overview of the orders of magnitude
involved; the values and sketches should only be considered as
indicative!

While many more parameters are involved, it is useful to restrain our
discussion to a two-dimensional space. Indeed the detection efficiency
of most of the observing techniques can be discussed in terms of the
orbital separation (or, equivalently, of the orbital period) and of
the mass- or flux-ratio of the components. For a given evolutionary stage, 
the mass-ratio can directly be related the flux ratio and  we will therefore 
assume a direct equivalence between these two values. This simplified approach assumes that
observations with sufficient time-sampling are available, and knowingly
neglects the second-order effects of eccentricity and orbital
inclination on the detection probabilities.

{\underline{\it Mass-ratio ($q=M_2/M_1$):}} In principle, the range of
possible mass-ratios spans equal-mass binaries ($q=1.0$) to a system
with a massive star with a light companion ($q<<1$). For example, an
O5\,$+$\,M8 system would have a mass ratio of only $q\sim0.002$. Of
course, a companion with such a low mass would be very hard to detect,
but the absence of observational clues does not preclude their
existence.  There are other observational issues, such as the
likelihood that low-mass companions are still in the pre-main sequence
phase -- observations at longer wavelengths could provide crucial
information in this scenario.  The range of flux-ratios that require
scrutiny can reach up to 10$^5$, providing a significant observational
challenge.

{\underline{\it Separations (d):}} An estimate of the minimal
separation can be adopted as the distance at which two main-sequence
stars would enter a contact phase. For typical O- and early B-type
primaries, this corresponds to rough separations of 20~R$_\odot$ or
0.1~AU, equating to periods of 1-2 days depending of the system
mass. The outer separation boundary is more of a grey zone that 
depends on both the system environment and on the timescale involved. In
this context, we consider two arguments. The first makes the
distinction between {\it hard} and {\it soft} binary systems, i.e.,
between systems that have a large likelihood of surviving a three-body
interaction, versus systems that
will be easily disrupted.  \citet{Heg75} defined {\it hard} binaries
as systems in which the binding energy ($E_b$) is larger than the kinetic
energy ($E_k$) brought about by an encounter :
\begin{equation}
|E_b|>E_k(encounter)=\frac{<m><v^2>}{2},
\end{equation} where $<m>$ and $<v^2>$ are the typical mass and velocity dispersions of  stars in a given cluster. Following \citet{PZMMG10} and adopting an effective cluster radius of 1~pc and cluster masses in the range 2.5$\times$10$^3$ to 10$^5$~M$_\odot$, one estimates the maximum separation of {\it hard} binaries to be in the range of 10$^3$ to several 10$^4$~A.U.

A second more qualitative argument emphasized by \citet{MAp10} points
out that massive stars have short life-times. One could therefore limit
the parameter space to orbital periods of 10$^5$ to 10$^6$~yr as only
these systems would accomplish a significant number of orbits during
their life-time. Following the third Kepler law, this also
corresponds to typical separations of several 10$^4$ AU.
 Interestingly, this means that most of the
massive binaries are hard binaries, that will be difficult to disrupt
over their life-time. The observed maximum range of separations considered
here is in line with the statement of \citet{Abt88}
that the more massive stars can sustain companions up to
several 10$^4$ AU\ or more.\\

\begin{figure}[t]
% \vspace*{-2.0 cm}
\begin{center}
 \includegraphics[width=13cm]{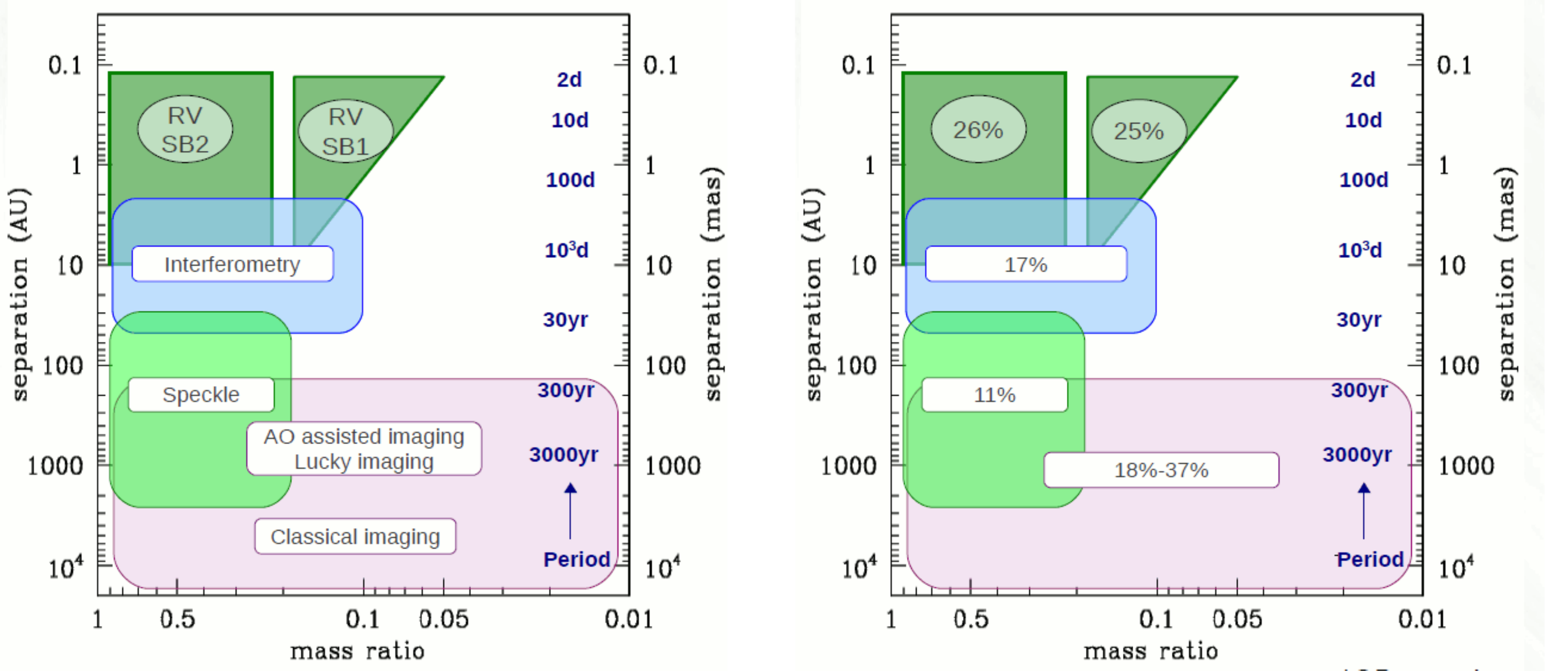}
% \vspace*{-1.0 cm}
\caption{Left-hand panel: typical parameter space for massive
binaries. A primary of 40~M$_\odot$ at a distance of 1~kpc has
been assumed to construct this sketch. The relevant regions for
various detection techniques have been overlaid. Right-hand panel:
measured multiplicity in those parts of parameter space (see text for
details).} \label{fig: paramspace}
\end{center}
\end{figure}

{\underline{\it Observational techniques:}} Investigating such a
large parameter space requires a combination of techniques (Fig.~\ref{fig: paramspace}), each
characterized by their own sensitivities and observational biases.
Short-period close binaries are probed efficiently through
spectroscopy, while very wide binaries, with angular separations
larger than a couple of arcseconds can be detected by classical, high-contrast
imaging. Enhanced imaging techniques such as adaptive optics
(AO) and lucky imaging can provide about an order of magnitude in
terms of closer separation and can also reach large flux
contrasts. In principle, the gap between the spectroscopic and imaging
regimes can be bridged with speckle interferometry, and
ground-based and space interferometry. Speckle interferometry has
the potential for large surveys but, to date, its applications have been limited
to flux ratios of about ten \citep{MHG09}. Space and ground-based
interferometry can reach separations of milliarcsecond scales, at flux
ratios of up to 100, but are much more costly to operate and
no large survey has yet been attempted.

Combining these various methods allows us in principle to explore the full range
of separations for massive binaries out to a distance of
$\approx5$~kpc. In practise, these techniques are not equally
sensitive and do not offer the same detection probability in their
respective regions of parameter space. For example, spectroscopy is
very efficient for short-period binaries, with periods of up to a couple
of years. The detection probability however decreases dramatically for
long-period systems \citep[see, e.g., Fig.~2 of ][]{EBB10}, in part
due to the reduced radial velocity (RV) signal and also due to the
longer timescales involved. Moreover, eccentric systems are
harder to detect due the narrower window
(sometimes less than a tenth of the orbital cycle) during which the RV
variations are concentrated. Imaging techniques (classical, lucky, or
AO-corrected) share a common bias in which the achievable contrast
varies as a function of the separation \citep[see e.g., Fig.~2 of][]{MAp10}.

Detailed comprehension of the limitations of each technique and of
their observational bias is of prime importance in order to 
 retrieve the global multiplicity properties of massive star
populations.

%%%%%%%%%%%%%%%%%%%%%%%%%%%%%%%%%%%%%%%%%%%%%%%%%%%%%%%%%%%%%%%%%%%%%%%%%%%%%%%%%%%%%%%%%%%%%%%%%%%%%%%%%
%%%%%%%%%%%%%%%%%%%%%%%%%%%%%%%%%%%%%%%%%%%%%%%%%%%%%%%%%%%%%%%%%%%%%%%%%%%%%%%%%%%%%%%%%%%%%%%%%%%%%%%%%
 
\section{The multiplicity fraction of O-type stars}\label{sect: fbin}

\subsection{Spectroscopic binary fraction in various separation regimes}

The right-hand panel of Fig.~\ref{fig: paramspace} gives an overview
of the results from recent surveys, including the minimum multiplicity
fraction obtained in each part of parameter space from the
relevant technique:

{\underline{\it Spectroscopy:}} The most comprehensive overview of the
spectroscopic binary (SB) fraction is provided by \citet{MHG09}. Based
on a review of the literature covering more than 300 O-type objects,
these authors found over half of the sample to be part of a SB
system. The systems are separated, almost equally, into single- (SB1)
and double-lined (SB2) systems.

{\underline{\it Speckle interferometry:}} In the same paper,
\citet{MHG09} provide speckle observations of 385 O-type stars, thus covering
almost all of the targets in the Galactic O star catalog
\citep{MAWG04}. 11\%\ of the objects in the \citeauthor{MHG09} sample are
found to have speckle companions.

{\underline{\it Enhanced imaging techniques:}} At larger separations,
AO-corrected and lucky imaging surveys (respectively, \citet{TBR08}
-- 138 O stars -- and \citet{MAp10} -- 128 O stars) found that 37\%\
of the O stars are part of wide multiple systems. These two studies are
mostly limited to the northern hemisphere and are thus missing
some of the richer massive star clusters and associations in the
southern sky. Part of this gap is filled by the AO campaigns of
\citet{DSE01} and \citet{SMG10} on, respectively, NGC~6611 and
Tr~14. Both studies revealed a lower multiplicity fraction of 18\%\
for their sample of OB stars. Yet, (part of) this difference results
from the fact that these two regions are dense clusters. In these
environments, disentangling the true pairs from chance alignment with
stars in the same clusters becomes more challenging and only a smaller
separation range can be investigated reliably. Interestingly, both
\citeauthor{DSE01} and \citeauthor{SMG10} concluded that
OB stars have more companions than lower mass-stars.

{\underline{\it Interferometry:}} As mentioned earlier, interferometry
is less suitable for surveys. To the best of our knowledge, only one
homogeneous survey has been attempted so far. \citet{NWW04} targeted a
limited sample of 23 O-type stars in the Carina region with the {\em Hubble
Space Telescope} fine guidance sensor, resolving close-by companions for
four stars.\\

Combining information from these various ranges, a minimum
multiplicity fraction close to 70\%\ for the population of Galactic
O-type stars is reached \citep{MHG09}. Given the detection limits of
these campaigns, there is ample scope for the true multiplicity
fraction to be even larger.   
 
Despite the quality of the observations collected so far, improvements are still
needed in each of the ranges covered by the various observing techniques described above:
\begin{enumerate}
\item[-] Homogeneous AO and lucky imaging campaigns have been mostly limited to the northern sky. Extending such work to the rich and dense clusters and star-formation regions of the southern hemisphere is highly desirable,
\item[-] Higher flux contrasts are needed in the 10-100~mas separation regime. Techniques such as sparse-aperture masking coupled with AO could, in principle, bring some improvements,
\item[-] The separation range 5-100~AU remains almost unexplored,
\item[-] About half  the known and suspected SBs lack an orbital solution. As a consequence, the distribution of the the orbital parameters remains largely uncertain (see also Section~\ref{sect: CDF}). 
 \end{enumerate}

%%%%%%%%%%%%%%%%%%%%%%%%%%%%%%%%%%%%%%%%%%%%%%%%%%%%%%%%%%%%%%%%%%%%%%%%%%%%%%%%%%%%%%%%%%%%%%%%%%%%%%%%%
\begin{table}[t]
  \begin{center}
  \caption{Overview of the spectroscopic binary fraction in clusters.}
  \label{tab: clusters}
 {\scriptsize
  \begin{tabular}{|l c c c |l c c c|}\hline 
{\bf Object} & {\bf\# O stars} & {\bf Binary fraction$^a$} & {\bf Ref} & {\bf Object} & {\bf\# O stars} & {\bf Binary fraction$^a$} & {\bf Ref.} \\ \hline
\multicolumn{4}{|c|}{\bf Nearby clusters} & \multicolumn{4}{c|}{\bf Distant/extragalactic clusters} \\
NGC 6611 &  9 & 0.44 & 1 & West 1    & 20 & 0.30 &  9 \\                                                     
NGC 6231 & 16 & 0.63 & 2 & 30 Dor     & 54 & 0.45 & 10 \\                                                     
IC 2944  & 14 & 0.53 & 3 & NGC346     & 19 & 0.21 & 11 \\                                                     
Tr 16    & 24 & 0.48 & 4 & N11        & 44 & 0.43 & 11 \\                                                     
IC 1805  &  8 & 0.38 & 5 & NGC2004    &  4 & 0.25 & 11 \\                                                     
IC 1848  &  5 & 0.40 & 5 & NGC 330    &  6 & 0.00 & 11 \\ 
NGC 2244 &  6 & 0.17 & 6 &            &    &      &    \\
Tr 14    &  6 & 0.00 & 7 &  \multicolumn{4}{c |}{\bf Milky Way O star population} \\ 
Col 228  & 15 & 0.33 & 8 & \begin{tabular}{c}Clusters \& \\  OB associations\end{tabular} & 305 & 0.57 & 12 \\
 \hline
  \end{tabular}
  }
 \end{center}
\vspace{1mm}
 \scriptsize{ {\it Notes:} $^a$The quoted binary fraction is a lower
 limit as each new detection will increase it.  \\ 
{\it References:} 1. \citet{SGE09},
 2. \citet{SGN08}, 3. \citet{SJG10}, 4. Literature review,
 5. \citet{HGB06}, 6. \citet{MNR09}, 7. \citet{PGH93}, \citet{GML98}, 8. Sana et al. (in prep.), 9. \citet{RCN09}, 10. \citet{BTT09}, 11. \citet{ELS06},
 12. \citet{MHG09} \\ }

\end{table}
%%%%%%%%%%%%%%%%%%%%%%%%%%%%%%%%%%%%%%%%%%%%%%%%%%%%%%%%%%%%%%%%%%%%%%%%%%%%%%%%%%%%%%%%%%%%%%%%%%%%%%%%%

\subsection{Spectroscopic binary fraction in clusters}

\citet{MHG09} investigated the dependence of the SB fraction on environment by
comparing stars from clusters and associations with runaway and field
stars, finding that the first category harbours many more binaries and
multiple systems. This picture is mostly consistent with an ejection
scenario for the field/runaway stars in which most of the multiple systems
would be disrupted.  In this section, we take a different approach and
look for differences in the multiplicity fraction of various
clusters. Several authors have indeed proposed the SB fraction to be
related to the cluster density \citep[e.g.,][]{PGH93, GM01}.

\begin{figure}[t]
% \vspace*{-2.0 cm}
\begin{center}
 \includegraphics[width=12cm,height=6cm]{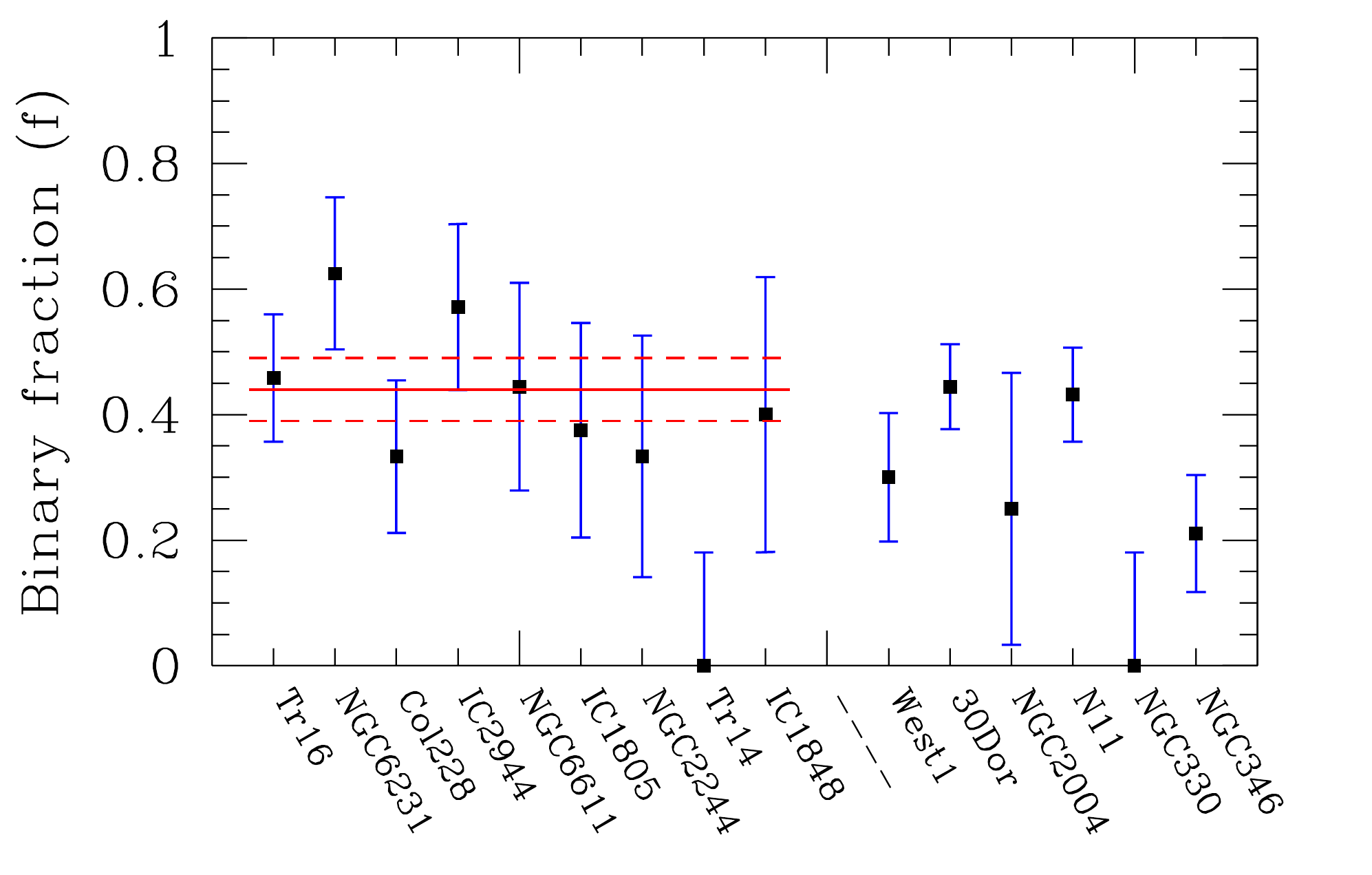}
% \vspace*{-1.0 cm}
 \caption{Spectroscopic binary fraction of nearby (left) and distant/extragalactic (right) clusters. The plain line and dashed lines indicate the average fraction and 1$\sigma$ dispersion computed from the nearby cluster sample. }
 \label{fig: clusters}
\end{center}
\end{figure}

To support our discussion, Table~\ref{tab: clusters} summarizes the SB
fraction of O-star rich clusters (i.e., clusters with at least five
O-type stars), with Fig.~\ref{fig: clusters} giving a graphical
comparison of the SB fractions in the various samples. Focusing on the
qualitatively homogeneous sample formed by the nearby clusters, we
calculate an average binary fraction of $f=0.44\pm0.05$. 
While some deviations are observed
around this average value, each can be explained by statistical
fluctuations. Even the extreme case of Tr~14, with no known
spectroscopic companions to its six O-type stars, is not statistically
significant. For instance, the probability to have six single stars,
drawn from an underlying binomial distribution with a multiplicity
fraction of $f=0.44$ is 3\%. 
Assuming that parent population is the same, the chance of obtaining zero binaries in any one of our nine clusters (given the size of their respective O star population) is 13\%, such that we cannot reject the null hypothesis.
Of course, the fact that Tr~14 is
the densest and possibly the youngest of the nine clusters in our
sample is intriguing. 

%%%%%%%%%%%%%%%%%%%%%%%%%%%%%%%%%%%%%%%%%%%%%%%%%%%%%%%%%%%%%%%%%%%%%%%%%%%%%%%%%%%%%%%%%%%%%%%%%%%%%%%%%
%%%%%%%%%%%%%%%%%%%%%%%%%%%%%%%%%%%%%%%%%%%%%%%%%%%%%%%%%%%%%%%%%%%%%%%%%%%%%%%%%%%%%%%%%%%%%%%%%%%%%%%%

The multiplicity properties from distant and extragalactic clusters
are less constrained and should be considered as lower limits, in part
because some of these works have a limited baseline and/or a limited
number of epochs. Aside from the case of NGC~330, there is again no
fundamental disagreement with the results from the nearby cluster
sample. With no companions seen for six O-type stars, the NGC~330
sample is similar, in terms of size and binary fraction, to
Tr~14. Sample size effects could be invoked (as for Tr~14), but the
fact that the much larger population of B-type stars in NGC~330 also show a
depleted binary population \citep{ELS06} is appealing. Interestingly,
NGC~330 is an older region with a very low surface density, in strong contrast with
the properties of Tr~14.

In summary, while some variations of the binary fraction might occur
in peculiar situations, the null hypothesis of a common parent
distribution cannot be rejected given the current data set. Adopting a uniform
binary fraction of $f\approx0.44$ is thus the most relevant
description of the current data. As a direct consequence of this
result, one can however reject with a very high confidence the null hypothesis
that all O stars are spectroscopic binaries.

%%%%%%%%%%%%%%%%%%%%%%%%%%%%%%%%%%%%%%%%%%%%%%%%%%%%%%%%%%%%%%%%%%%%%%%%%%%%%%%%%%%%%%%%%%%%%%%%%%%%%%%%%
\begin{table}[t]
  \begin{center}
  \caption{Overview of the two O star samples used to derive the distributions of the orbital parameters. The first part of the table indicates the number of O  stars, the number of O-type binaries and the binary fraction of the two samples. The second part of the table provides the number and the fraction of systems with constraints on their periods, mass-ratios and eccentricities. }
  \label{tab: sample}
 {\scriptsize
  \begin{tabular}{|l c c c |}\hline 
{\bf } & {\bf\# Galactic O stars} && {\bf Nearby rich clusters}  \\ \hline
\# O stars            & 305 && 82 \\
\# binaries           & 173 && 38 \\
Binary fraction       & 0.57 && 0.46 \\
\hline
\# periods            & 102 (59\%)  && 33 (87\%) \\
\# mass-ratios        &  76 (44\%)  && 29 (76\%) \\
\# eccentricities     &  86 (50\%)  && 30 (79\%) \\
 \hline
  \end{tabular}
  }
 \end{center}
\vspace{1mm}
 \scriptsize{
 {\it Note:}
  The sample of nearby clusters is formed by IC 1805, IC1848, IC 2944, NGC6231, NGC 6611 and Tr16.\\
}
\end{table}
%%%%%%%%%%%%%%%%%%%%%%%%%%%%%%%%%%%%%%%%%%%%%%%%%%%%%%%%%%%%%%%%%%%%%%%%%%%%%%%%%%%%%%%%%%%%%%%%%%%%%%%%%

\section{Distributions of the orbital parameters of spectroscopic binaries}\label{sect: CDF}
This section provides an overview of our current knowledge of the orbital parameter distributions for O-type spectroscopic binaries. In doing so, it is useful to define two samples (Table~\ref{tab: sample}):
\begin{enumerate}
\item[-] {\underline{\it The Galactic O-star sample:}} mostly based on the sample of \citet{MHG09}.
While \citeauthor{MHG09} only concentrate on the multiplicity aspect,
we perform our own literature review to search for estimates of
periods, mass-ratios and eccentricities. When no orbital solution was
available, we estimated the mass-ratios of SB2 systems by adopting
typical masses for the components as a function of their spectral
classification \citep{MSH05}. Compared to the review of \citet{MHG09}, we also
include information that became available in the last two years, as
well as preliminary results from our work.
\item[-] {\underline{\it The nearby O-star rich clusters:}} a subsample of
the Galactic O-star sample, focusing on the O-star rich clusters
within $\approx3$~kpc. These clusters have been more thoroughly
studied so that the scope for observational biases is more limited.
\end{enumerate}

The binary fraction of the two samples appear to be different, with
the Galactic O-star sample displaying more binaries. A possible
explanation for this is provided by \citet{GM01}, who noted that the O
stars in poor clusters (i.e., clusters with only one or two O-type stars)
were almost all multiple. These clusters are not included in our
second sample, which may pull the binary fraction to lower values.\

\begin{figure}[t]
% \vspace*{-2.0 cm}
\begin{center}
 \includegraphics[width=10cm]{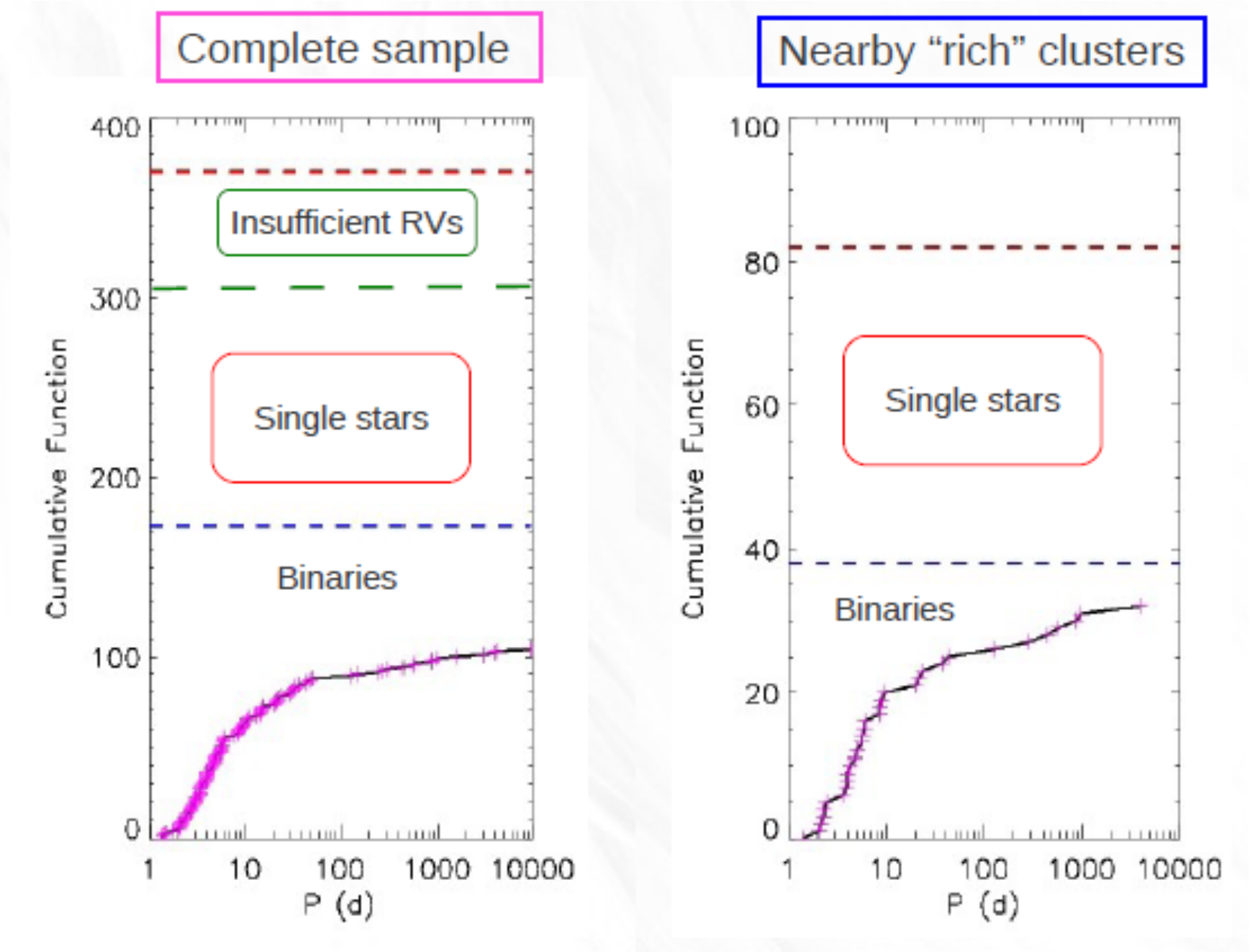} 
% \vspace*{-1.0 cm}
 \caption{Cumulative number function of orbital periods for the
 complete sample (left) and for the nearby cluster sample
 (right). This plot aims to give a graphical impression of the potential 
 biases affecting the two samples. Normalised cumulative
 distribution functions for systems with solutions are given in Fig.~\ref{fig: CDF}.}
 \label{fig: Pcdf_sample}
\end{center}
\end{figure}

While the Galactic O-star sample is the most comprehensive, only about
50\%\ of the binaries have constraints on their orbital solution
(Fig.~\ref{fig: Pcdf_sample}), leaving a lot of room for observational
biases. For example, the orbital solutions are more difficult to obtain for
long-period high eccentricity systems. There might thus be an uneven
representation of various parameter ranges in the observed
distribution functions. The situation is much improved for the cluster
sample, as almost 80\%\ of the systems have proper orbital solutions
and 87\%\ have estimates of the orbital period. We therefore argue
that the distributions derived from the cluster sample are much less 
affected by observational biases.
 In the following, we will compare the parameter
distributions built from the two samples to one another and to
analytical distributions commonly used to represent the properties of
the massive star binary population.\\

%%%%%%%%%%%%%%%%%%%%%%%%%%%%%%%%%%%%%%%%%%%%%%%%%%%%%%%%%%%%%%%%%%%%%%%%%%%%%%%%%%%%%%%%%%%%%%%%%%%%%%%%%
\begin{figure}[t]
% \vspace*{-2.0 cm}
\begin{center}
\includegraphics[width=13cm]{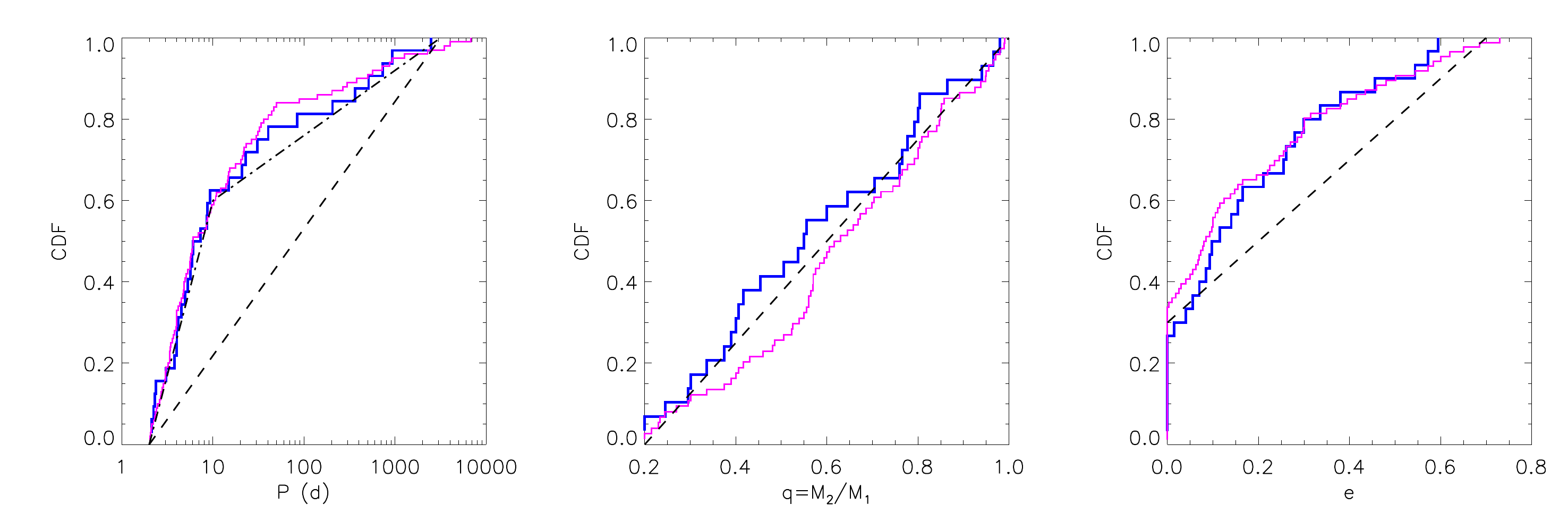}
% \vspace*{-1.0 cm}
\caption{Cumulative distribution functions (CDFs) of the periods
($P$), mass-ratios ($q$) and eccentricities ($e$). The plain thin/magenta and thick/blue lines indicate the CDFs of, the Galactic O-star sample and
the nearby cluster sample, respectively. Left-hand panel: the dashed line shows an \"{O}pik
Law over this range of periods, while the dot-dashed line indicates the
alternative law given by Eq.~\ref{eq: CDFp2}. Middle panel: the
dashed line indicates a uniform distribution in the considered
range. Right-hand panel: the dashed line indicates a uniform distribution
for $e>0$.}  \label{fig: CDF}
\end{center}
\end{figure}

%%%%%%%%%%%%%%%%%%%%%%%%%%%%%%%%%%%%%%%%%%%%%%%%%%%%%%%%%%%%%%%%%%%%%%%%%%%%%%%%%%%%%%%%%%%%%%%%%%%%%%%%
{\underline{\it Period:}}  Fig.~\ref{fig: Pcdf_sample} provides an
overview of the respective samples with the cumulative number
distributions of the orbital periods. It shows that the period
distribution function obtained from the cluster sample is almost fully
constrained, but that uncertainties could still affect the
Galactic sample. However, the cumulative distribution functions (CDFs)
are mostly in agreement (Fig.~\ref{fig: CDF}, left-hand panel). Both CDFs
show an overabundance of short periods, with 50 to 60\%\ of the systems
having a period shorter than 10 days. Consequently, the CDF of observed
periods in the spectroscopic regime can not be represented by the
traditional \"{O}pik Law\footnote{\"{O}pik's Law states that
the distribution of separations is flat in logarithmic space. The
corresponding period distribution should be flat in $\log P$ as
well.}. As already suggested by \citet{SGN08}, a much better
representation of the period CDF is provided by a bi-uniform
distribution in $\log P$ (which one could consider a `broken'
\"{O}pik Law) such that:
\begin{equation}
CDF(P) = \left\{ 
 \begin{array}{l l}
\frac{F_\mathrm{break} \bigl( \log P-\log P_\mathrm{min} \bigl)}{\log P_\mathrm{break}-\log P_\mathrm{min}}, &   \mathrm{for}\ \log P_\mathrm{min} \le \log P \le P_\mathrm{break} \\
\\
F_\mathrm{break}+\frac{ \bigl(1-F_\mathrm{break} \bigl) \bigl(\log P-\log P_\mathrm{break}\bigl)}{\log P_\mathrm{max}-\log P_\mathrm{break}} , &  \mathrm{for}\  P_\mathrm{break}< \log P \le \log P_\mathrm{max} \\
\end{array}\right.
\label{eq: CDFp}
\end{equation}
where $P$ is expressed in days. Adopting a break-point at
$P_\mathrm{break}\approx10$~d, with upper and lower limits of $\log
P/\mathrm=0.3$ and 3.5\,d and considering that the binaries are evenly
spread in the short and long period regimes (i.e.,
$F_\mathrm{break}\approx0.5$), Eq.~\ref{eq: CDFp} becomes:
\begin{equation}
CDF(P) =\left\{ 
 \begin{array}{l l}
\frac{5}{7} \log P-\frac{10.5}{7}, & \quad \mathrm{for}\ 0.3 \le \log P \le 1.0 \\
\\
\frac{1}{5} \log P-\frac{3}{10}, & \quad\mathrm{for}\ 1.0 < \log P \le 3.5  \\
\end{array}\right.
\label{eq: CDFp2}
\end{equation}

%%%%%%%%%%%%%%%%%%%%%%%%%%%%%%%%%%%%%%%%%%%%%%%%%%%%%%%%%%%%%%%%%%%%%%%%%%%%%%%%%%%%%%%%%%%%%%%%%%%%%%%%%
%%%%%%%%%%%%%%%%%%%%%%%%%%%%%%%%%%%%%%%%%%%%%%%%%%%%%%%%%%%%%%%%%%%%%%%%%%%%%%%%%%%%%%%%%%%%%%%%%%%%%%%%

Eqs. \ref{eq: CDFp} and \ref{eq: CDFp2} give an empirical description
of the CDF of the observed periods. The latter should still be
corrected for the detection probability (mostly affecting longer
periods) and for the systems lacking orbital solutions (also more
likely to affect the longer-period regime). The exact location of the
lower and upper limits and of the `break' still needs to be more
tightly constrained. That said, the general behaviour and the
overabundance of short-period spectroscopic binaries appear clear.

{\underline{\it Mass-ratio:}} The CDFs of the mass-ratios
(Fig.~\ref{fig: CDF}, middle panel) are well reproduced by a uniform
distribution in the range 0.2\,$<$\,$q$\,$<$\,1.0. The Galactic O-star sample
shows slightly fewer systems with $q<0.6$; this can be (partly) 
explained by observational biases as the detection of the secondary
signature for systems with large mass differences (i.e., large flux contrasts)
requires very high-quality data that are not always available for the Galactic sample. SB1
binaries represent about 20-25\%\ of the cluster sample. For these
stars, one cannot directly estimate the mass-ratio. However, we note
that the fraction of SB1 is roughly compatible with an extension of
the uniform CDF towards $q<0.2$; testing this statement will require
detailed simulations.

As a direct consequence of a uniform mass-ratio CDF, the presence of a
twin population with $q>0.95$ proposed by \citet{PiS06} can be rejected.
Another implication resides in the fact that massive binaries cannot
be formed by random pairing from a Salpeter/Kroupa IMF. Our results
rather suggest the presence of a mechanism that favors the creation of
O\,$+$\,OB binaries. Such a mechnaism could find part of its origin in the
early dynamical evolution, where companion exchanges favor the capture
of more and more massive secondaries. It could also trace a
particular formation mechanism \citep{ZiY07}. 

{\underline{\it Eccentricity:}} The CDF of the eccentricities
(Fig.\ref{fig: CDF}, right-hand panel) is characterized by an overabundance
of circular and low eccentricity systems.  Indeed, 25-30\%\ of the systems
displayed a circular orbit, while another 30\%\ have $e<0.2$. This
behaviour contradicts the expected properties of a purely thermal
binary population, which can be qualitatively explained by the large
fraction of short-period systems for which tidal dissipation will tend
to circularize the orbit.

An analytical description of the observed CDF for eccentric systems
can be provided through $CDF(e>0) \propto e^{0.5}$ in the range
0.0\,$<$\,$e$\,$<$\,0.8. However, as 20\%\ of the cluster sample and
50\%\ of the Galactic sample are lacking robust eccentricities and as
biases are most likely to affect larger eccentricities, we cannot
consider this relation as definite. That said, one would expect that
$CDF(e)$ will remain overabundant towards low eccentricity systems.

\section{Summary}\label{sect: ccl}

We have attempted to provide an overview of our current knowledge of
the important multiplicity properties of massive stars. We described
some of the physical processes and observational biases that lead
to binaries behaving differently compared to single stars.
We then briefly described the observational parameter space that one
needs to explore to investigate massive binaries, and we discussed the
challenges of probing it homogeneously.  Despite these difficulties,
it is now well established that the vast majority of O-type stars are part
of a multiple system. The typical separation between the multiple
components covers at least 4 order of magnitudes.

At least 45-55\%\ of the O star population in clusters and OB
associations is comprised by spectroscopic binaries, with a lower
fraction found for field and runaway stars \citep{MHG09}.  Here we
have investigated possible variations of the multiplicity fraction
among clusters with a rich O star population. While room for small
variations remains due to our limited sample and due to the small O
star population of some clusters, the binary fraction can mostly be
considered as uniform with a value close to 44\%. Given the current data set,
one can hardly argue that the multiplicity fraction is significantly correlated 
with the cluster density (at least not in the range covered in our sample)
. While density can still play a role, for example, to explain
the difference observed between O-star rich and O-star poor
clusters, its impact among rich clusters remain questionable in
light of the current data. It is well accepted that most 
O-type stars are part of a multiple systems, but a similar statement does not
hold when limiting ourselves to spectroscopic companions. Given 
the observed SB binary fraction and the sample sizes, it is
unlikely that the underlying fraction of SBs is larger than 70-75\%.

Finally, we have constructed CDFs for the periods, mass-ratios and
eccentricities for two samples of massive binaries. The Galactic
O-star sample is more extensive but has been studied less
homogeneously. The second sample, based on the O star binary
population in six rich nearby open clusters, is more homogeneous and
is less susceptible to detection biases. There are some differences 
in the CDFs of the two samples (see Fig.~\ref{fig:
CDF}), but two-sided Kolmogorov-Smirnoff tests do not reveal
statistically significant deviations. These differences can be
qualitatively understood in terms of different observational effects.
 Currently, the observed CDFs for $P$, $q$
and $e$ of spectroscopic O-type binaries can be analytically described
by the following functions:
\begin{enumerate}
\item[-] {\em Periods:} a broken \"{O}pik Law with a break point at $P\sim10$~days,
\item[-] {\em Mass-ratios:} a uniform distribution down to $q=0.2$, potentially extending in the SB1 domain (i.e., for $q<0.2$),
\item[-] {\em Eccentricities:} 25-30\%\ of the characterised systems have circular orbits. $CDF(e>0)$ shows a 
square-root dependance with $e$, but detailed considerations of bias are lacking at present.
\end{enumerate}
A quantitative analysis of the effects of the detection limit and of
other observational biases would be highly desirable (although not
trivial) in order to: (i) assess the completness and the exactness of
the observed CDFs; (ii) retrieve the underlying distributions. \\

In conclusion, significant progress has been made in the past two
decades but uncertainties on the exact multiplicity properties of
massive stars remain numerous. In particular, an homogeneous
exploration of the parameter space, the distribution function of the
orbital parameters and the impact of the environment on the
multiplicity properties are likely the areas in which observational
progresses are the most crucially needed.  
Fortunately, numerous projects are currently underway which aim at
improving our knowledge of these aspects. It is our hope to have drawn
attention to the importance of a proper understanding of the detection
limits and of the observational biases that affect each survey. These
are necessary information to consider in order to glue all the
pieces together toward a global view of the massive star properties
across the full reach of parameter space and in different
environements.

\section*{Acknowledgments}
The authors warmly thank the organizers for their invitation and for their flexibility. The authors also wished to express their thanks to M. De Becker, A.\ de Koter, S. de Mink, M. Gieles, E. Gosset, P. Massey and S. Portegies Zwart for useful discussion in the preparation and redaction of this review.

\bibliographystyle{aa.bst}
\bibliography{/home/hsana/PAPERS/literature}

\end{document}